# Effects of the linear polarization of polariton condensates in their propagation in co-directional couplers

Elena Rozas, Alexey Yulin, Johannes Beierlein, Sebastian Klembt, Sven Höfling, Oleg A. Egorov, Ulf Peschel, Ivan A. Shelykh, Manuel Gundin, Ignacio Robles-López, M. Dolores Martín, Luis Viña

Elena Rozas
Departamento de Física de Materiales, Universidad Autónoma de Madrid, 28049 Madrid, Spain; Instituto Nicolás Cabrera, Universidad Autónoma de Madrid, 28049 Madrid, Spain.

Alexey Yulin
Faculty of Physics and Engineering ITMO University, St. Petersburg 197101, Russia.

Johannes Beierlein
Technische Physik, Wilhelm-Conrad- Röntgen-Research Center for Complex Material Systems, and Würzburg-Dresden Cluster of Excellence ct.qmat, Universität Würzburg, D-97074 Würzburg, Germany.

Sebastian Klembt
Technische Physik, Wilhelm-Conrad- Röntgen-Research Center for Complex Material Systems, and Würzburg-Dresden Cluster of Excellence ct.qmat, Universität Würzburg, D-97074 Würzburg, Germany.

Sven Höfling
Technische Physik, Wilhelm-Conrad-Röntgen- Research Center for Complex Material Systems, and Würzburg-Dresden Cluster of Excellence ct.qmat, Universität Würzburg, D-97074 Würzburg, Germany; SUPA, School of Physics and Astronomy, University of St. Andrews, St. Andrews KY16 9SS, United Kingdom.

Oleg A. Egorov
Institute of Condensed Matter Theory and Optics Friedrich-Schiller-University Jena, D-07743 Jena, Germany.

Ulf Peschel
Institute of Condensed Matter Theory and Optics Friedrich-Schiller-University Jena, D-07743 Jena, Germany.

I. A. Shelykh
Faculty of Physics and Engineering ITMO University, St. Petersburg 197101, Russia; Science Institute. University of Iceland. Reykjavik IS-107, Iceland.

Manuel Gundin
Departamento de Física de Materiales, Universidad Autónoma de Madrid, 28049 Madrid, Spain.

Ignacio Robles-López
Departamento de Física de Materiales, Universidad Autónoma de Madrid, 28049 Madrid, Spain; Instituto Nicolás Cabrera, Universidad Autónoma de Madrid, 28049 Madrid, Spain.

M. D. Martín
Departamento de Física de Materiales, Universidad Autónoma de Madrid, 28049 Madrid, Spain; Instituto Nicolás Cabrera, Universidad Autónoma de Madrid, 28049 Madrid, Spain.
E-mail: dolores.martin@uam.es

L. Viña
Departamento de Física de Materiales, Universidad Autónoma de Madrid, 28049 Madrid, Spain; Instituto Nicolás Cabrera, Universidad Autónoma de Madrid, 28049 Madrid, Spain; Instituto de Física de la Materia Condensada, Universidad Autónoma de Madrid, 28049 Madrid, Spain.

ABSTRACT: We report on the linear polarization of polariton condensates in a co-directional coupler that allows evanescent coupling between adjacent waveguides. During the condensate's formation, polaritons usually acquire a randomly oriented polarization, however, our results reveal a preferential orientation of the linear polarization along the waveguide propagation path. Furthermore, we observe polarization-dependent intensity oscillations in the output terminal of the coupler, and we identify the mode beating between the linear-polarized eigenmodes as the origin of these oscillations. Our findings provide an insight into the control of the polarization of polariton condensates, paving the way for the development of spin-based polaritonic architectures where condensates propagate over macroscopic distances.

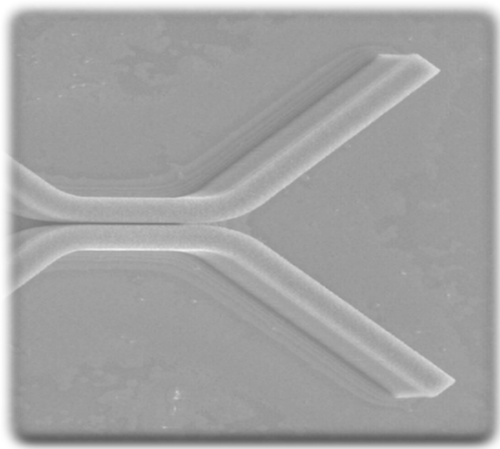
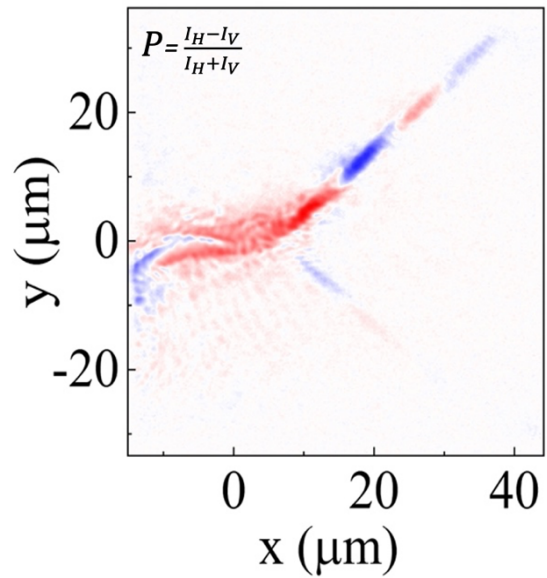

1. Introduction

Microcavity exciton polaritons have been, in the latest years, the subject of numerous investigations given their exceptional properties [1, 2]. These properties, emerging from the strong coupling between their constituents, excitons and photons, allow polaritons to behave as bosonic particles. Their short lifetime, typically of the order of ps, is comparable or even smaller than their respective thermalization times, so in general these particles do not reach thermal equilibrium. However, a condensation similar to a Bose-Einstein one is observed when the particle density is increased [3-7]. Their very low effective mass ($\sim 10^{-4}$ $m_e$, being $m_e$ the free electron mass) can lead to condensation even at room temperature. This has been indeed the case in transition metal dichalcogenides, organic semiconductor materials and lead halide perovskites structures, where the enhancement of the exciton binding energy, characteristic of these compounds, has enabled the observation of strong light-matter coupling [8,9], polariton lasing and condensation [10-15] at room temperature. The ease of use and the flexibility offered by exciton polaritons unwrapped a variety of new proposed devices such as polariton interferometers [16], logic gates [17,18] or transistors [19-21]. Additionally, three dimensional polariton confinement has been achieved in microcavity pillars [22-26]. Using two dimensional lattices of these micropillars, it is possible to emulate graphene and its remarkable properties [27-29]. All these devices are built using refined lithographic techniques that ensure the polariton confinement along several directions. In the one-dimensional case, only a well-defined longitudinal path along which polariton condensates can travel remains [18, 19, 30, 31].

In the present work, we will focus on semiconductor microcavity couplers. Such optical directional couplers are formed by parallel optical waveguides, closely spaced, so that energy exchange can occur between them [32]. The coupled power, limited by the mode's overlap in the coupler arms, is determined by the separation between the waveguides, the wavelength, the evanescence of the modes and the interaction length. These devices have been proved to be essential for splitting and combining light in photonic systems and have been used widely in the silicon-on insulator platform [33]. Quantum photonic waveguide circuits based on GaAs/Ga$_{1-x}$Al$_x$As heterostructures have been demonstrated for the manipulation of quantum states of light [34]. These devices have been also exploited for guiding surface plasmon polaritons [35, 36] and exciton polaritons [37-39]. More recently, we have reported on different on-chip routing devices: a counter-directional coupler [40] and a co-directional coupler for condensates of exciton-polaritons, studying the peculiarities of the polariton propagation [41] and how this is affected by the waveguides' energetic landscape [42]. A relevant factor is the spin state of the condensates after polariton's relaxation processes leading to their condensation [43, 44]. Moreover, a spontaneous build-up of the linear polarization of the emitted light above the polariton condensation threshold has been reported both theoretically [45-47] and experimentally [4, 48-51]. The orientation of the polarization plane of the emission is pinned to a crystallographic axis of the microcavity [50, 51]. This effect has been effusively observed for trapped polaritons using different trapping mechanism such as photonic disorder [52], stress [5] or annular optical confinement [53]. The study and control of the polarization state of polariton condensates has opened new possibilities of designing and improving spin-based devices [54-57]. Wire-shaped microcavities are particularly interesting in this respect because, due to their reduced symmetry, each polariton mode shows a polarization splitting into two modes polarized along and perpendicular to the wire axis [58].

Here, we theoretically and experimentally study the linear polarization of the emission of propagating polariton condensates in polaritonic co-directional couplers. Our results demonstrate a coupling between the adjacent waveguides that is not strongly dependent on polarization. However, we encounter striking polarization-dependent emission oscillations at the output terminal of the coupler. For a given set of perpendicular polarizations we find a phase shift between the oscillation's patterns. To better understand our experimental results, a dissipative Gross-Pitaevskii model is used to describe the polarization dynamics in the device.

2. Experimental details

The sample used in this work is a $\lambda/2$ cavity with top (bottom) distributed Bragg reflectors consisting of 23 (27) pairs of alternating layers of $Al_{0.2}Ga_{0.8}As/AlAs$. Three stacks of 4 GaAs quantum wells of 7 nm of nominal width are grown at the antinodes of the electromagnetic field inside the cavity. Low power measurements reveal a Q-factor of ~ 5000 and a Rabi splitting of 13.9 meV. The experiments reported here are performed in a region of the sample with a photon-exciton detuning $\delta \approx -17$ meV. The sample has been grown by molecular beam epitaxy and processed by reactive ion etching down to the QWs [41], creating a pattern of adjacent waveguides where, length (*L*), width (*w*) and separation (*d*), have been varied. Figure 1(a) shows a typical field of couplers, with different coupling lengths, formed by doubly bent waveguides with input and output terminals rotated ±45º from the longitudinal direction; the geometrical parameters are specified in Fig. 1(b). The part of the device where both waveguides remain parallel along the x-direction is dubbed coupling region: a few pairs of mirrors left in the region between the two arms enable the evanescent photonic coupling of polaritons between the guides [41]. For the experiments reported here, the dimensions of the selected device are *L* = 10 µm, *w* = 2 µm and *d* = 0.2 µm. The choice of these parameters allows the coupling of a large fraction of polaritons between the arms of the coupler. In our experiments, we non-resonantly pump the input terminal of the coupler with 2-ps pulses from a Ti:Al$_2$O$_3$ laser working at 1.664 eV, focusing the beam to a 4.5 µm diameter spot, with a microscope objective (NA = 0.40, f = 4 mm), impinging normally to the sample surface. The photoluminescence (PL) is collected through the same objective while the sample is kept at 12 K in a cold-finger, He flow cryostat, and detected with a CCD camera attached to a 0.5 m focal length imaging spectrometer. We ensure that the polariton condensation threshold (12 kW/cm$^2$) has been exceeded and that condensates propagate along the entire device pumping with a power density of 26 kW/cm$^2$.

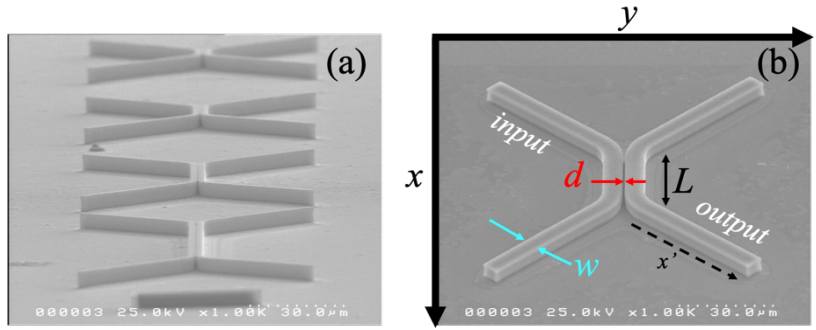

**Figure. 1.** (a) SEM image of a field of directional polariton couplers with different coupling region length (*L*). (b) SEM image of a directional polariton coupler indicating several parameters: coupling length (*L*= 20 µm), waveguide width (*w*= 6 µm) and waveguide separation (*d* = 0.6 µm). Input, output terminals and the coordinates axis are shown, corresponding to the nomenclature used in the text: *x* (*y*) parallel (perpendicular) to the main axis of the waveguides in the coupling region and *x'* along the axes of the input and output terminals at ∓45° with respect to *x* and *y*, respectively

## 3.- Theoretical framework

To study the dynamics of exciton-polaritons theoretically we adopt a well-known model describing the coherent polaritons by two complex order parameter functions $\Psi_{r,l}$ for right ($r$) and left ($l$) handed circularly polarized polaritons [55, 59]. The coherent polaritons interact with baths of incoherent excitons having different spins. The excitons are characterized by their density $n_{r,l}$. The whole set of equations can be written as:

$$i\hbar \partial_t \Psi_{r,l} = \frac{i\hbar}{2}(Rn_{r,l} - \gamma_p)\Psi_{r,l} +$$
$$\left(V(x,y) + G|\Psi_{r,l}|^2 + \widetilde{G}|\Psi_{l,r}|^2 + G_R n_{r,l} + \widetilde{G_R} n_{l,r}\right)\Psi_{r,l} - \frac{\hbar^2}{2m_{eff}}\nabla^2 \Psi_{r,l} + B(\partial_x \pm i\partial_y)^2 \Psi_{l,r} + A_{r,l}(x,y,t) \quad (1)$$

$$\partial_t n_{r,l} = -\left(\Gamma_e + R|\Psi_{r,l}|^2\right)n_{r,l} + P_{r,l}(x,y,t). \quad (2)$$

In these equations, $R$ is the coupling parameter between the polaritons and the reservoirs of excitons, $\gamma_p$ is the coordinate dependent losses of the coherent polaritons. We assume that the polariton waveguides are formed by microstructuring creating a coordinate-dependent effective potential, $V$, for the polaritons. It is also considered that the microstructuring affects the transparency of the Bragg mirrors, thus the losses experienced by polaritons become larger outside the waveguides. G ($\widetilde{G}$) and $G_R$ ($\widetilde{G_R}$) denote nonlinear corrections –blue shift– to the effective potential due to interactions between polaritons and between polaritons and incoherent excitons of the same (orthogonal) polarization, respectively. $m_{eff} = \frac{2m_\perp m_\parallel}{m_\perp + m_\parallel}$ is the reduced effective mass of polaritons, with $m_\perp$ and $m_\parallel$ the transverse and longitudinal polariton masses, while $B = \frac{m_\perp - m_\parallel}{m_\perp + m_\parallel}$ defines the strength of TE-TM splitting (spin-orbit coupling). $\Gamma_e$ accounts for the linear losses in the exciton subsystem, and $P_{r,l}$ is the intensity of the optical pump creating the exciton baths. In the experiments reported here, an incoherent pump has been used in order to create a bath of excitons that are responsible for setting the polaritons in motion [30]. However, polaritons can also be excited resonantly by coherent light. This kind of excitation provides a simpler scenario in numerical calculations to control the properties of the polaritons; we use it in our numerical simulations to clarify some aspects of polariton dynamics when required. In the Eqs. (1)-(2) this pump is accounted by the driving force $A_{r,l}(x,y,t)$, the last term on the right-hand side of Eq. (1).

## 4.- Propagation of polaritons along the waveguides

The propagation distance of these condensates following the input waveguide axis ($x'$) is displayed in Figure 2. The zero position on the horizontal axis corresponds to the excitation spot. Polariton condensates propagate ballistically away from the excitation spot in the lower input terminal after being generated, as readily seen in the inset, which also shows the coupling to the upper arm. The solid black and dot-dashed red lines depict the experimental PL intensity and the numerical simulation (see below), respectively, for polaritons moving towards the coupling region. Close to the origin, the experimental PL is much smaller than that of the simulations due to energy filtering performed by the spectrometer. Due to the finite polariton lifetime, the polariton population decays exponentially as it moves away from the excitation area. The simulations describe well the experimental results up to $x' \sim 20$ µm, where the intensity droops when reaching the waveguide bent.

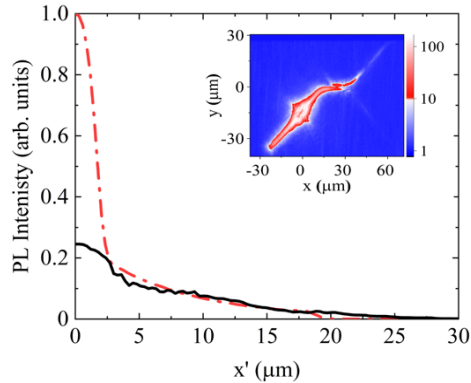

**Figure. 2.** Polariton propagation along the longitudinal axis of the input arm of a polariton coupler. The $x'$ origin is placed at the excitation spot. The photoluminescence (PL) intensity (black line) is plotted along the axis of the input terminal, $x'$, in the region ($0 < x < 21$, $-15 < y < 0$) together with the results of the simulation (red line). The inset shows the experimental map of the coupler emission from which the black curve was extracted.

Now we describe the condensates propagation in the coupling region and in the output terminal. Since the PL intensity in these regions is substantially lower than in the input terminal, we have spatially filtered the emission, so that the PL from the input terminal is removed. The excitation beam is vertically polarized ($\theta_i = 90º$), i.e., parallel to the y axis. The PL is analyzed using a linear polarizer at different angles, ranging from $\theta_d = 0º$ (i.e., horizontal polarization) up to 180º in steps of 10º. For simplicity, only a summary of the PL for selected $\theta_d$ is shown in Figure 3. Polariton condensates are generated in the bottom-left input terminal; when they arrive to the coupling region, -5 < x < 5 μm, a large fraction of the population is conveyed from the bottom to the top arm. After the coupling, polaritons continue propagating throughout the top arm until the edge of the waveguide at the output terminal, while only a minor fraction of the population remains in the bottom arm. By increasing either the length (*L*) or the spacing (*d*) between the arms, the fraction of coupled polaritons can be controlled [41]. Drastic intensity variations along the device are observed when the polarization of the emission is analyzed. We find a considerably large intensity when the polarization is analyzed at 0°. By contrast, a remarkable intensity reduction is observed around and above 30°. A further increase of $\theta_d$ results in a slow PL recovery for $\theta_d \gtrsim 90°$.

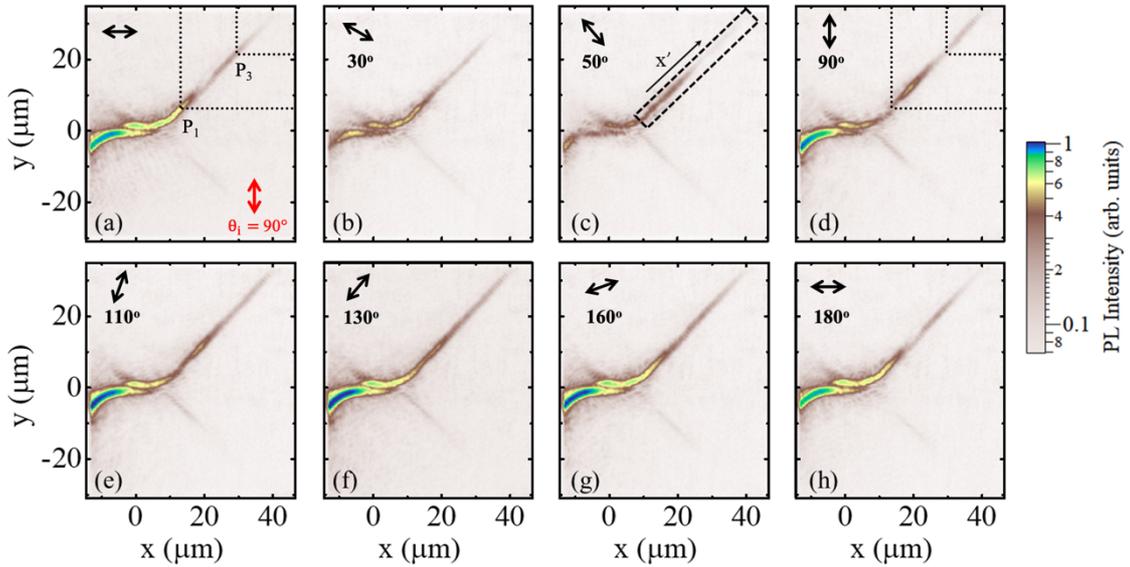

**Figure. 3.** PL of polariton condensates propagating along a coupler. The emission is analyzed at different linear polarizations (black arrows) ranging from $\theta_d$= 0° (a) to 180° (h). The excitation at the input terminal (not shown) is performed with a vertically polarized laser beam [denoted by the green arrow in (a)]. The output terminal in which polarization-dependent oscillations are visible for some angles $\theta_d$ ′s is indicated by a black dashed rectangle in (c); the axis along this terminal is defined as x′. The positions labeled as $P_1$ (where i =1,3) mark the maximum amplitude of the aforementioned oscillations for $\theta_d$= 0°. The PL emissions are filtered at an energy of 1.583 eV and depicted in a normalized logarithmic false-color scale. A power density of 26 kW/cm$^{-2}$ has been used for the measurements.

Furthermore, a conspicuous additional effect is observed at the output terminal. The PL on this terminal displays two local maxima at $P_1$ = (13, 7) and $P_3$ = (30, 22) μm when the polarization is filtered at 0°. This is in stark contrast with the emission observed at $\theta_d$ = 90°, in which local minima appear at the same sets of coordinates. These intensity oscillations in the PL do not exist for intermediate polarizations (see for instance 50° and 130º): the emission shows just an exponential decay with propagation distance along the output terminal. This behavior is independent of the excitation laser polarization as borne out by our experiments, since the non-resonant excitation conditions in our case guarantee the erasing of polariton's spin memory during the relaxation processes (See also Figure S1).

Let us now discuss how the theoretical model introduced above describes the effects observed in the experiment. We performed numerical simulations of the waveguides with the experimental geometrical parameters and with a depth and width of the effective polariton potential providing the width of the fundamental mode (full width at half maximum of the intensity) to be very close to 1 μm: the width of the mode in the modeled waveguide is equal to the width of the fundamental mode of an infinitely deep rectangular potential of 2 μm wide. In the numerical simulations, we use a random noise of low intensity as initial conditions and excite the system by an incoherent pump creating a bath of excitons, which eventually create the polariton condensates. The results of the numerical simulations shown in Figure 4 attest a good qualitative agreement with the experiments: they reproduce the observation that the polariton condensate is preferentially polarized along the direction of the waveguide.

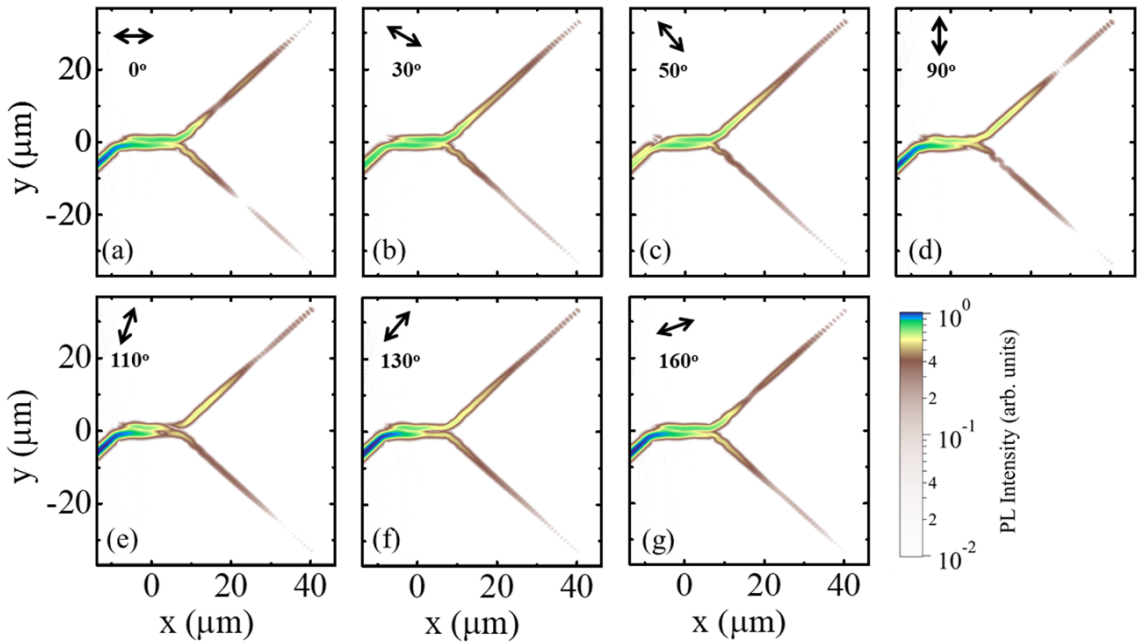

**Figure. 4.** Simulations of the polariton distribution along a coupler. The emission is analyzed at different linear polarizations (black arrows) ranging from $\theta_d$= 0° (a) to 160° (g). The excitation is performed with a linearly polarized laser beam at the input terminal (not shown). The density of polaritons is depicted in a normalized logarithmic false-color scale.

Additionally, the numerical modeling replicates qualitatively the experiments bearing out the tunneling of polaritons from the lower to the upper waveguide in the coupling region. Figure 5 presents a comparison between the experimental results [panel (a)] and the calculations [panel (b)]: the normalized PL intensity is plotted as a function of y, the transverse to the axis of the coupling region coordinate, for a position close to the entrance/exit (-3.2/+3.2 µm) of that region with a solid/dash-dotted line (for $\theta_d = 0$). Note that y = 0 marks the center of the gap between the arms, therefore, the signal from y < 0 and y > 0 arises from the pumped and coupled arm, respectively. It is apparent that at the entrance, a larger polariton population is present in the pumped arm of the coupler (y < 0) than that in the coupled arm (y > 0), both in the experiments and the simulations. This situation is reversed towards the exit, where the population becomes larger in the coupled arm.

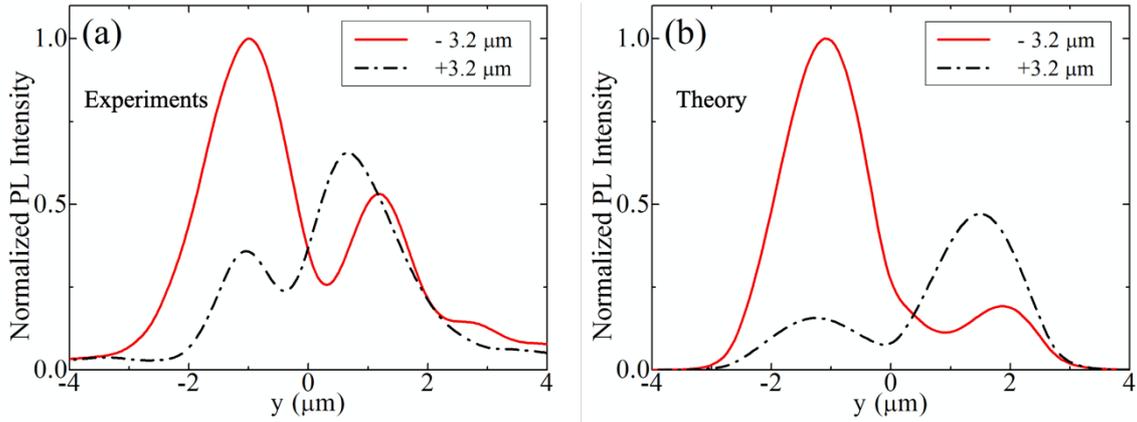

**Figure. 5.** (a) The solid/dash-dotted line depicts the experimental normalized PL intensity vs. y, the transverse to the axis of the coupling region coordinate, at a position close to the entrance/exit of the coupling region, -3.2 µm /+3.2 µm , for $\theta_d = 0°$. (b) Corresponding simulated polariton densities at the same positions of the coupler as those shown in panel (a).

The amount of polaritons transferred to the upper arm depends on the height and the width of the potential separating the upper and the lower arms as well as on the dissipation rate of the propagating polaritons, which experimentally is greatly influenced by sample inhomogeneities. The lack of a full quantitative agreement between the theoretical and the experimental polariton transfer can be attributed to the not perfect fitting of the experimental polariton dissipation rate and the height of the effective confinement potential used in the simulations.

It should also be stated that, both in the experiments and theory, the population transfer is not strongly polarization dependent. In the experiments a 65(±4) % of the polaritons are transferred in the coupling region regardless of their polarization.

Finally, and most importantly for the present work, the polarization beating seen in the output terminal of the upper waveguide is unmistakably obtained in the simulations. This effect originates from the TE-TM splitting of the polaritons: it is theoretically discussed in detail in the next Section and in the Supporting Information (SI).

5.- Polariton's polarization dynamics.

To quantitatively describe the polarized PL, we focus now on two regions of the coupler: the coupling region and the output terminal. In the coupling region, the PL intensity is found experimentally to be maximum for $\theta_d = 0º$ (= 180º), i.e., when the polarization is horizontal, parallel to the longitudinal axis in this region. On the contrary, the intensity drops with increasing $\theta_d$ reaching its minimum at ~ 45º. A further increase of $\theta_d$ leads to a partial recovery of the emission, resulting in a PL for $\theta_d = 90º$ (vertical polarization) ~ 50 % lower than that of $\theta_d = 0º$. These results reveal a preferential orientation of the polarization along the longitudinal axis of the waveguide. For larger $\theta_d$, being the polarized emission direction closer to the orientation of the pumped (input) terminal, the PL intensity in the coupled region becomes high again as a consequence of a larger population with this polarization in the input terminal. This is readily seen in the polarization maps compiled in Figure 6. Panel (a)/(c) shows the experimental/simulated degree of polarization, defined as $P=\frac{I_H-I_V}{I_H+I_V}$, where $I_H$ ($I_V$) is the PL intensity for $\theta_d = 0º$ (90º), and demonstrates the preferential polarization orientation in the coupling region, where a positive degree of polarization (red coded values) is obtained both in the experiments and the simulations. A positive degree of polarization in this region is also obtained in the diagonal basis, as depicted in panel (b)/(d) for $P=\frac{I_D-I_A}{I_D+I_A}$ obtained in the experiments/simulations; in this case $I_D$ ($I_A$) is the PL intensity for $\theta_d = 50º$ (130º).

We focus now on the polarization-dependent oscillations at the output terminal. These oscillations are clearly seen as a sign change of *P* (going alternatively to red- and blue-coded values), both in the experiments and the simulations, in panels (a) and (c) of Fig.

6, that show the polarization degree in the H\V basis. However, they vanish when the polarization is analyzed in the D\A polarization basis [see panels (b) and (d)]. For a better understanding of this effect, PL profiles have been extracted along the x′ direction of the terminal, in the region marked with a dashed line in Fig. 3(c). Figure 7 summarizes the experimental (a) and simulated (b) emission profiles for different polarizations between $\theta_d = 0º$ and 180º in steps of 10º. The zero position denotes the beginning of the output terminal while the end of the structure is located at x′ ~ 45 μm. A transition between different patterns can be distinguished. For $\theta_d$ close to 0º and 180º, beatings separated by ~ 24 μm are observed, with maxima at $x_1'$ ~ 4 μm and $x_3'$ ~ 28 μm. In contrast, for $\theta_d$ close to 90º, a weak signal is obtained at $x_1'$ being the maximum now at $x_2'$ ~ 16 μm, i.e., out of phase from $x_1'$ by half the beating distance (12 μm).

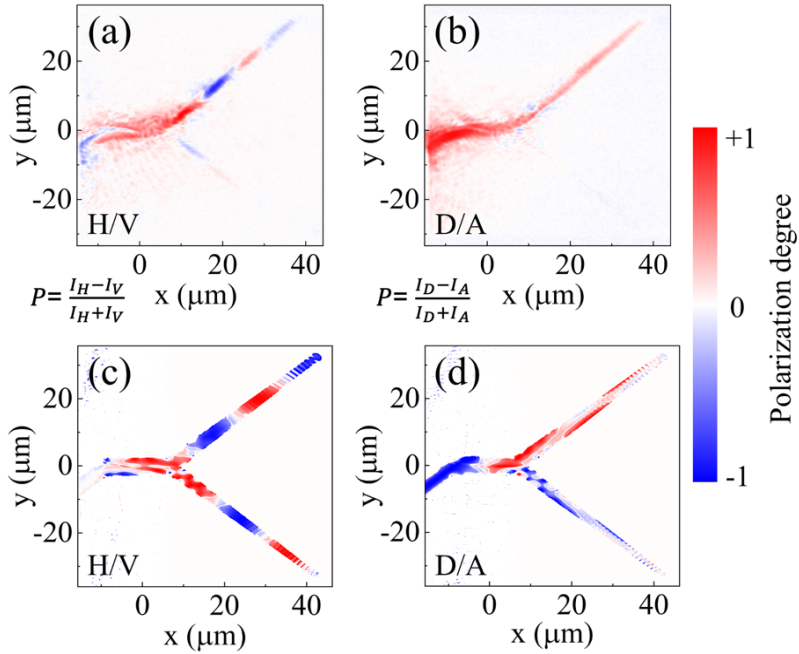

**Figure. 6.** (a) Spatial maps of the polariton polarization degree, $P=\frac{I_H-I_V}{I_H+I_V}$, for $\theta_d = 0°$ (H) and $\theta_d = 90°$ (V) polarizations of the emission in the coupler. (b) Corresponding maps for $\theta_d = 50°$ (D) and $\theta_d = 130°$ (A) rendering $P=\frac{I_D-I_A}{I_D+I_A}$. (c) and (d) show the analogous degrees of polarization obtained from the simulated polariton densities for the same analyzer angles as those presented in (a) and (b), respectively. The degree of polarization is plotted in a false color scale with red (blue) corresponding to positive (negative) values.

We can explain these polarization beatings by considering the TE-TM splitting in the waveguide and therefore taking into account the different group velocities for each polarization. In the framework of the mathematical model introduced above, the splitting lifts the degeneracy between the eigenmode polarized along (TE) and that polarized perpendicularly (TM) to the waveguide axis [60]. At a given frequency, these two modes have different wavevectors and therefore, if both modes are excited, interference fringes are obtained when the polarization of the propagating polaritons is a linear combination of those of the eigenmodes. The beating period is given by $L_{beating} = \frac{2\pi}{|k_\parallel - k_\perp|}$, where $k_\parallel$ and $k_\perp$ are the wavevectors of the TE and TM modes, respectively. This is the case along the output terminal for $\theta_d = 0°$ and 90°, which are linear combinations of the TM (45°) and TE (135°) modes in this part of the coupler, where clear beatings are observed both in the experiments and simulations. However, the beats are absent for $\theta_d \sim 45°$ and $\sim 135°$.

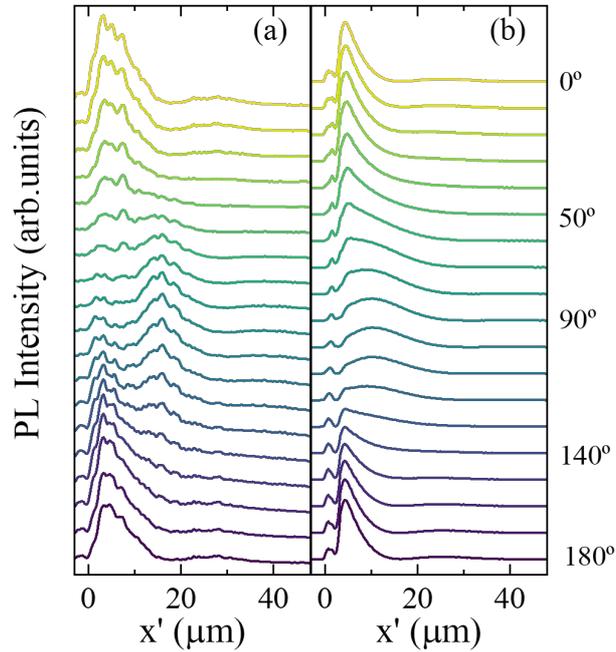

**Figure. 7.** (a) PL intensity, integrated along the transverse coordinate in the output terminal of the coupler, vs. distance, measured along the longitudinal axis at this terminal, for different analyzer angles, ranging from θd = 0° (top) to 180° (bottom) in steps of 10°. (b) Corresponding simulated polariton densities for the same θd's.

We would like to mention here that PL intensity fringes having a short spatial period are also clearly seen at the ends of the waveguides. These short-period oscillations are visible since polariton coherence is preserved during their propagation [61, 62]. They originate from the reflection of the polaritons at the waveguide end resulting in the formation of counter-propagating polariton waves that interfere with the incoming ones [36].

6.- Conclusions

In summary, we have evidenced the rich phenomenology of polariton propagation in co-directional couplers when the population is analyzed into its linearly polarized components. A detailed analysis of the PL has been accomplished by mapping the polarized-dependent emission at the condensate's energy. Two different sections of the coupler have been studied in detail, the coupling region and the output terminal. In the former region, it has been shown a transfer of polariton population from the pumped- to the coupled-arm of the device that is not strongly polarization dependent. In the latter region, on the contrary, polarization-dependent oscillations emerging from the TE-TM splitting of the fundamental modes of the waveguide have been found. In these co-directional couplers, based on waveguide microcavities, which provide a polarization dependent mode spectrum, the polariton spin gains crucial importance. Our work paves the way for the use polariton co-directional couplers taking advantage of the changes of the polariton's polarization state during their propagation. Further effects of the geometry of the couplers on the spin of the polariton condensates, such as the influence of the bends on the circular polarization, are currently under investigation.

**Supporting information.** Movie of polarization degree of propagating polaritons. Additional experiments on polarization beatings at the output terminal. Simulations of the polariton dynamics. Excitation by either a coherent or incoherent pump.

**Funding**

This work has been partly supported by the Spanish MINECO Grant No. MAT2017-83722-R. AY and IAS were financially supported by the Ministry of Science and Higher Education of the Russian Federation through Megagrant number 14.Y26.31.0015 and goszadanie no. 2019-1246, IAS acknowledges also the support from the Icelandic research fund, grant No. 163082-051. The Würzburg and Jena group acknowledges financial support within the DFG Projects No. PE 523/18-1 and No. KL3124/2-1. The Würzburg group acknowledges financial support by the German Research Foundation (DFG) under Germany's Excellence Strategy–EXC2147 "ct.qmat" (Project No. 390858490) and is grateful for support by the state of Bavaria.

**ACKNOWLEDGEMENTS**: We thank C. Schneider and H. Suchomel for sample growth and useful discussions.

**References**

[1] Weisbuch, C.; Nishioka, M.; Ishikawa, A.; Arakawa, Y. Observation of the coupled exciton-photon mode splitting in a semiconductor quantum microcavity. *Phys. Rev. Lett*. **1992**, *69*, 3314-3317.

[2] Kavokin, A.; Baumberg, J.; Malpuech, G.; Laussy, F. *Microcavities*; Oxford science publications: Oxford, 2017.

[3] Dang, L.S.; Heger, D.; André, R.; Bœuf, F.; Romestain, R. Stimulation of polariton photoluminescence in semiconductor microcavity. *Phys. Rev. Lett.* **1998**, *81*, 3920-3923.

[4] Kasprzak, J.; Richard, M.; Kundermann, S.; Baas, A.; Jeambrun, P.; Keeling, J.; Marchetti, F.M.; Szymańska, M.H.; André, R.; Staehli, J.L.; Savona, V.; Littlewood, P.B.; Deveaud, B.; Dang, L. S. Bose–Einstein condensation of exciton polaritons. *Nature* **2006**, *443*, 409-414.

[5]     Balili, R.; Hartwell, V.; Snoke, D.; Pfeiffer, L.; West, K. Bose-Einstein condensation of microcavity polaritons in a trap. *Science* **2007**, *316*, 1007-1010.

[6]     Deng, H.; Haug, H.; Yamamoto, Y. Exciton-polariton Bose-Einstein condensation. *Rev. Mod. Phys.* **2010**, *82*, 1489-1537.

[7]     Byrnes, T.; Kim, N.; Yamamoto, Y. Exciton–polariton condensates. *Nat. Phys.* **2014**, *10*, 803-813.

[8]     Liu, X.; Galfsky, T.; Sun, Z.; Xia, F.; Lin, E.; Lee, Y.; Kéna-Cohen, S.; Menon, V.M. Strong light–matter coupling in two-dimensional atomic crystals. *Nat. Photon.* **2015**, *9*, 30-34.

[9]     Ardizzone, V.; de Marco, L.; de Giorgi, M.; Dominici, L.; Ballarini, D.; Sanvitto, D. Emerging 2D materials for room-temperature polaritonics. *Nanophotonics* **2019**, *8*, 1547-1558.

[10]    Kéna-Cohen, S.; Forrest, S.R. Room-temperature polariton lasing in an organic single-crystal microcavity. *Nat. Photon.* **2010**, *4*, 371-375.

[11]    Su, R.; Diederichs, C.; Wang, J.; Liew, T.C.H.; Zhao, J.; Liu, S.; Xu, W.; Chen, Z.; Xiong, Q. Room-temperature polariton lasing in all-inorganic perovskite nanoplatelets. *Nano Lett.* **2017**, *17*, 3982-3988.

[12]    Su, R.; Wang, J.; Zhao, J.; Xing, J.; Zhao, W.; Diederichs, C.; Liew, T.C.H.; Xiong, Q. Room temperature long-range coherent exciton polariton condensate flow in lead halide perovskites. *Sci. Adv.* **2018**, *4*, eaau0244.

[13]    Dusel, M.; Betzold, S.; Egorov, O.A.; Klembt, S.; Ohmer, J.; Fischer, U.; Höfling, S.; Schneider, C. Room temperature organic exciton–polariton condensate in a lattice. *Nat. Commun.* **2020**, *11*, 2863.

[14]    Shan, H.; Lackner, L.; Han, B.; Sedov, E.; Rupprecht, C.; Knopf, H.; Eilenberger, F.; Yumigeta, K.; Watanabe, K.; Taniguchi, T.; Klembt, S.; Höfling, S.; Kavokin, A.V.; Tongay, S.; Schneider, C.; Antón-Solanas, C. Coherent light emission of exciton-polaritons in an atomically thin crystal at room temperature. *arXiv:2103.10459 [cond-mat.mes-hall]* **2021**, na. Anton-Solanas, C.; Waldherr, M.; Klaas, M.; Suchomel, H.; Harder, T.H.; Cai, H.; Sedov, E.; Klembt, S.; Kavokin, A.V.; Tongay, S.; Watanabe, K.;

Taniguchi, T.; Höfling, S.; Schneider, C. Bosonic condensation of exciton–polaritons in an atomically thin crystal. *Nat. Mater*. **2021**, (doi: 10.1038/s41563-021-01000-8).

[15]    Su, R.; Ghosh, S.; Wang, J.; Liu, S.; Diederichs, C.; Liew, T.C.H.; Xiong, Q. Observation of exciton polariton condensation in a perovskite lattice at room temperature. *Nat. Phys*. **2020**, *16*, 301-306.

[16]    Sturm, C.; Tanese, D.; Nguyen, H.S.; Flayac, H.; Galopin, E.; Lemaître, A.; Sagnes, I.; Solnyshkov, D.; Amo, A.; Malpuech, G.; Bloch, J. All-optical phase modulation in a cavity-polariton Mach–Zehnder interferometer. *Nat. Commun*. **2014**, *5*, 3278.

[17]    Espinosa-Ortega, T.; Liew, T.C.H. A complete architecture of integrated photonic circuits based on AND and NOT logic gates of exciton-polaritons in semiconductor microcavities. *Phys. Rev. B* **2013**, *87*, 195305.

[18]    Antón, C.; Liew, T.C.H.; Cuadra, J.; Martín, M.D.; Eldridge, P.S.; Hatzopoulos, Z.; Stavrinidis, G.; Savvidis, P.G.; Viña, L. Quantum reflections and shunting of polariton condensate wave trains: Implementation of a logic AND gate. *Phys. Rev. B* **2013**, *88*, 245307.

[19]    Gao, T.; Eldridge, P.S.; Liew, T.C.H.; Tsintzos, S.I.; Stavrinidis, G.; Deligeorgis, G.; Hatzopoulos, Z.; Savvidis, P.G. Polariton condensate transistor switch. *Phys. Rev. B* **2012**, *85*, 235102.

[20]    Antón, C.; Liew, T.C.H.; Tosi, G.; Martín, M.D.; Gao, T.; Hatzopoulos, Z.; Eldridge, P.S.; Savvidis, P.G.; Viña, L. Dynamics of a polariton condensate transistor switch. *Appl. Phys. Lett*. **2012**, *101*, 261116.

[21]    Ballarini, D.; de Giorgi, M.; Cancellieri, E.; Houdré, R.; Giacobino, E.; Cingolani, R.; Bramati, A.; Gigli, G.; Sanvitto, D. All-optical polariton transistor. *Nat. Commun*. **2013**, *4*, 1778.

[22]    Bajoni, D.; Peter, E.; Senellart, P.; Smirr, J.L.; Sagnes, I.; Lemaître, A.; Bloch, J. Polariton parametric luminescence in a single micropillar. *Appl. Phys. Lett*. **2007**, *90*, 051107.

[23] Galbiati, M.; Ferrier, L.; Solnyshkov, D.D.; Tanese, D.; Wertz, E.; Amo, A.; Abbarchi, M.; Senellart, P.; Sagnes, I.; Lemaître, A.; Galopin, E.; Malpuech, G.; Bloch, J. Polariton condensation in photonic molecules. *Phys. Rev. Lett*. **2012**, *108*, 126403.

[24] Antón, C.; Solnyshkov, D.; Tosi, G.; Martín, M.D.; Hatzopoulos, Z.; Deligeorgis, G.; Savvidis, P.G.; Malpuech, G.; Viña, L. Ignition and formation dynamics of a polariton condensate on a semiconductor microcavity pillar. *Phys. Rev. B* **2014**, *90*, 155311.

[25] Schneider, C.; Gold, P.; Reitzenstein, S.; Höfling, S.; Kamp, M. Quantum dot micropillar cavities with quality factors exceeding 250,000. *Appl. Phys. B* **2016**, *122*, 19.

[26] Bajoni, D.; Senellart, P.; Wertz, E.; Sagnes, I.; Miard, A.; Lemaître, A; Bloch, J. Polariton laser using single micropillar GaAs−GaAlAs semiconductor cavities. Phys. Rev. Lett. **2008**, *100*, 047401.

[27] Jacqmin, T., Carusotto, I.; Sagnes, I.; Abbarchi, M.; Solnyshkov, D.; Malpuech, G.; Galopin, E.; Lemaître, A.; Bloch, J.; Amo A. Direct observation of Dirac cones and a flatband in a honeycomb lattice for polaritons. *Phys. Rev. Lett.* **2014**, *112*, 116402.

[28] Jamadi, O.; Rozas, E.; Salerno, G.; Milićević, M.; Ozawa, T.; Sagnes, I.; Lemaître, A.; Gratiet, L.L.; Harouri, A.; Carusotto, I.; Bloch, J.; Amo, A. Direct observation of photonic Landau levels and helical edge states in strained honeycomb lattices. *Light: Science & Applications* **2020**, *9*, 144.

[29] Klembt, S.; Harder, T.H.; Egorov, O.A.; Winkler, K.; Ge, R.; Bandres, M.A.; Emmerling, M.; Worschech, L.; Liew, T.C.H.; Segev, M.; Schneider, C; Höfling, S. Exciton-polariton topological insulator. *Nature* **2018**, *562*, 552-556.

[30] Wertz, E.; Ferrier, L.; Solnyshkov, D.D.; Johne, R.; Sanvitto, D.; Lemaître A.; Sagnes, I.; Grousson, R.; Kavokin, A.V.; Senellart, P.; Malpuech, G.; Bloch, J. Spontaneous formation and optical manipulation of extended polariton condensates. *Nat. Phys.* **2010**, *6*, 860-864.

[31] Wertz, E.; Amo, A.; Solnyshkov, D.D.; Ferrier, L.; Liew, T.C.H.; Sanvitto, D.; Senellart, P.; Sagnes, I.; Lemaître, A.; Kavokin, A.V.; Malpuech, G.; Bloch, J. Propagation and Amplification Dynamics of 1D Polariton Condensates. *Phys. Rev. Lett*. **2012**, *109***,** 216404.

[32] Ebeling, K.J. *Directional Couplers*, in *Integrated Optoelectronics*; Springer Verlag: Berlin, 1993.

[33] Lu, Z.; Yun, H.; Wang, Y.; Chen, Z.; Zhang, F.; Jaeger, N.A.F.; Chrostowski, L. Broadband silicon photonic directional coupler using asymmetric-waveguide based phase control. *Optics Express* **2015**, *23*, 3795-3808 and references therein.

[34] Wang, J.; Santamato, A.; Jiang, P.; Bonneau, D.; Engin, E.; Silverstone, J.W.; Lermer, M.; Beetz, J.; Kamp, M.; Höfling, S.; Tanner, M.G.; Natarajan, C.M.; Hadfield, R.H.; Dorenbos, S.N.; Zwiller, V.; O'Brien, J.L.; Thompson, M.G. Gallium arsenide (GaAs) quantum photonic waveguide circuits. *Optics Commun*. **2014**, *327*, 49-55.

[35] Zhang, Y.; Xu, Y.; Tian, C.; Xu, Q.; Zhang, X.; Li, Y.; Zhang, X.; Han, J.; Zhang, W. Terahertz spoof surface-plasmon-polariton subwavelength waveguide. *Photonics Res.* **2018**, *6*, 18-23.

[36] Pan, M.-Y.; Lin, E.-H.; Wang, L.; Wie, P.-K. Spectral and mode properties of surface plasmon polariton waveguides studied by near-field excitation and leakage-mode radiation measurement. *Nanoscale Res. Lett*. **2014**, *9*, 430.

[37] Flayac, H.; Savenko, I. G. An exciton-polariton mediated all-optical router. *Appl. Phys. Lett*. **2013**, *103*, 201105.

[38] Marsault, F.; Nguyen, H.S.; Tanese, D.; Lemaître, A.; Galopin, E.; Sagnes, I.; Amo, A.; Bloch, J. Operation of a semiconductor microcavity under electric excitation. *Appl. Phys. Lett.* **2015**, *107*, 201115.

[39] Liao, L.; Ling, Y.; Luo, S.; Zhang, Z.; Wang, J.; Chen, Z. Propagation of a polariton condensate in a one-dimensional microwire at room temperature. *Appl. Phys. Express* **2019**, *12*, 052009.

[40] Klaas, M.; Beierlein, J.; Rozas, E.; Klembt, S.; Suchomel, H.; Harder, T.H.; Winkler, K.; Emmerling, M.; Flayac, H.; Martín, M.D.; Viña, L.; Höfling, S.; Schneider, C. Counter-directional polariton coupler. *Appl. Phys. Lett*. **2019**, *114*, 061102.

[41] Beierlein, J.; Rozas, E.; Egorov, O.A.; Klaas, M.; Yulin, A.; Suchomel, H.; Harder, T.H.; Emmerling, M.; Martín, M.D.; Shelykh, I.A.; Schneider, C.; Peschel, U.; Viña, L.; Hoffling, S.; Klembt, S. Propagative oscillations in codirectional polariton waveguide couplers. *Phys. Rev. Lett.* **2021**, *126,* 075302.

[42] Rozas, E.; Beierlein, J.; Yulin, A.; Klaas, M., Suchomel, H.; Egorov, O.; Shelykh, I.A.; Peschel, U.; Schneider, C.; Klembt, S.; Höfling, S.; Martín, M.D.; Viña, L. Impact

of the energetic landscape on polariton condensates' propagation along a coupler. *Adv. Opt. Mat.* **2020**, *8*, 2000650.

[43]     Martín, M.D.; Aichmayr, G.; Viña, L.; André, R. Polarization control of the nonlinear emission of semiconductor microcavities. *Phys. Rev. Lett*. **2002**, *89***,** 077402.

[44]     Lagoudakis, P.G.; Savvidis, P.G.; Baumberg, J.J.; Whittaker, D. M.; Eastham, P. R.; Skolnick, M.S.; Roberts, J.S. Stimulated spin dynamics of polaritons in semiconductor microcavities. *Phys. Rev. B* **2002**, *65*, 161310(R).

[45]     Laussy, F.P.; Shelykh, I.A.; Malpuech, G.; Kavokin, A. Effects of Bose-Einstein condensation of exciton polaritons in microcavities on the polarization of emitted light. *Phys. Rev. B* **2006**, *73*, 035315.

[46]     Malpuech, G.; Glazov, M.M.; Shelykh, I.A.; Bigenwald, P.; Kavokin, K.V. Electronic control of the polarization of light emitted by polariton lasers. *Appl. Phys. Lett.* **2006**, *88*, 111118.

[47]     del Valle, E.; Sanvitto, D.; Amo, A.; Laussy, F.P.; André, R.; Tejedor, C; Viña, L. Dynamics of the Formation and Decay of Coherence in a Polariton Condensate. *Phys. Rev. Lett.* **2009**, *103*, 096404.

[48]     Kasprzak, J.; André, R.; Dang, L. S.; Shelykh, I.A.; Kavokin, A.V.; Rubo, Y.G.; Kavokin, K.V.; Malpuech, G. Build up and pinning of linear polarization in the Bose condensates of exciton polaritons. *Phys. Rev. B* **2007**, *75*, 045326.

[49]     Baumberg, J.J.; Kavokin, A.V.; Christopoulos, S.; Grundy, A.J.D.; Butté, R.; Christmann, G.; Solnyshkov, D. D.; Malpuech, G.; Baldassarri Höger von Högersthal, G.; Feltin, E.; Carlin, J.F.; Grandjean, N. Spontaneous polarization buildup in a room-temperature polariton laser. *Phys. Rev. Lett.* **2008**, *101*, 136409.

[50]     Martín, M.D., Ballarini, D.; Amo, A.; Kłopotowsi, Ł.; Viña, L.; Kavokin, A.V.; André, R. Striking dynamics of II–VI microcavity polaritons after linearly polarized excitation. *Phys. Status Solidi C* **2005**, *2*, 3880-3883.

[51]     Kłopotowski, Ł.; Martín, M.D.; Amo, A.; Viña, L.; Shelykh, I.A.; Glazov, M.M.; Malpuech, G.; Kavokin, A.V.; André, R. Optical anisotropy and pinning of the linear polarization of light in semiconductor microcavities. *Solid State Commun.* **2006**, *139*, 511-515.

[52] Sanvitto, D.; Amo, A.; Viña, L.; André, R.; Solnyshkov, D.; Malpuech, G. Exciton-polariton condensation in a natural two-dimensional trap. Phys. Rev. B **2009**, *80*, 045301.

[53] Askitopoulos, A.; Nalitov, A.V.; Sedov, E. S.; Pickup, L.; Cherotchenko, E.D.; Hatzopoulos, Z.; Savvidis, P.G.; Kavokin, A.V.; Lagoudakis, P. G. All-optical quantum fluid spin beam splitter. *Phys. Rev. B* **2018**, *97*, 235303.

[54] Ballarini, D.; Amo, A.; Viña, L.; Sanvitto, D.; Skolnick, M.S.; Roberts, J.S. Transition from the strong- to the weak-coupling regime in semiconductor microcavities: Polarization dependence. *Appl. Phys. Lett*. **2007**, *90*, 201905.

[55] Kammann, E.; Liew, T.C.H.; Ohadi, H.; Cilibrizzi, P.; Tsotsis, P.; Hatzopoulos, Z.; Savvidis, P. G.; Kavokin, A.V.; Lagoudakis, P. G. Nonlinear optical spin Hall effect and long-range spin transport in polariton lasers. *Phys. Rev. Lett.* **2012**, *109*, 036404.

[56] Antón, C.; Morina, S.; Gao, T.; Eldridge, P.S.; Liew, T.C.H.; Martín, M.D.; Hatzopoulos, Z.; Savvidis, P.G.; Shelykh, I.A.; Viña, L. Optical control of spin textures in quasi-one-dimensional polariton condensates. *Phys. Rev. B* **2015**, *91*, 075305.

[57] Gao, T.; Antón, C.; Liew, T.C.H., Martín, M.D.; Hatzopoulos, Z.; Viña, L.; Eldridge P.S.; Savvidis P.G. Spin selective filtering of polariton condensate flow. *Appl. Phys. Lett.* **2015**, *107*, 011106.

[58] Dasbach, G.; Diederichs, C.; Tignon, J.; Ciuti, C.; Roussignol, Ph.; Delalande, C.; Bayer, M.; Forchel, A. Polarization selective polariton oscillation in quasi-onedimensional microcavities. *Phys. Status Solidi C* **2005**, *2*, 779-782.

[59] Wouters, M.; Carusotto, I. Excitations in a Nonequilibrium Bose-Einstein Condensate of Exciton Polaritons. *Phys. Rev. Lett.* **2007**, *99*, 140402.

[60] Shelykh, I.A.; Nalitov, A.V.; Iorsh, I.V. Optical analog of Rashba spin-orbit interaction in asymmetric polariton waveguides. *Phys. Rev. B* **2018**, *98*, 155428.

[61] Weeber, J.C.; Krenn, J. R.; Dereux, A.; Lamprecht, B.; Lacroute, Y.; Goudonnet, J.P. Near-field observation of surface plasmon polariton propagation on thin metal stripes. *Phys. Rev. B* **2001**, *64*, 045411.

[62] Antón, C.; Liew, T.C.H.; Tosi, G.; Martín, M. D.; Gao, T.; Hatzopoulos, Z.; Eldridge, P. S.; Savvidis, P.G.; Viña, L. Energy relaxation of exciton-polariton condensates in quasi-one-dimensional microcavities. *Phys. Rev. B* **2013**, *88*, 035313.

**Supplementary information: Effects of the linear polarization of polariton condensates in their propagation in direction couplers**

*Elena Rozas[1,2], Alexey Yulin[3], Johannes Beierlein[4], Sebastian Klembt[4], Sven Höfling[4,5], Oleg A. Egorov[6], Ulf Peschel[6], Ivan A. Shelykh[3,7], Manuel Gundin[1], Ignacio Robles-López[1,2], M. Dolores Martín[1,2]\* and Luis Viña[1,2,8]*

[1]Departamento de Física de Materiales, Universidad Autónoma de Madrid, 28049 Madrid, Spain

[2]Instituto Nicolás Cabrera, Universidad Autónoma de Madrid, 28049 Madrid, Spain.

[3]Faculty of Physics and Engineering ITMO University, St. Petersburg 197101, Russia.

[4]Technische Physik, Wilhelm-Conrad- Röntgen-Research Center for Complex Material Systems, and Würzburg-Dresden Cluster of Excellence ct.qmat, Universität Würzburg, D-97074 Würzburg, Germany.

[5]SUPA, School of Physics and Astronomy, University of St. Andrews, St. Andrews KY16 9SS, United Kingdom.

[6]Institute of Condensed Matter Theory and Optics Friedrich-Schiller-University Jena, D-07743 Jena, Germany.

[7]Science Institute. University of Iceland. Reykjavik IS-107, Iceland.

[8]Instituto de Física de la Materia Condensada, Universidad Autónoma de Madrid, 28049 Madrid, Spain.

## I. Polarization degree of propagating polaritons.

Movie showing maps of propagating polaritons for their emission analyzed into its linearly polarized components.

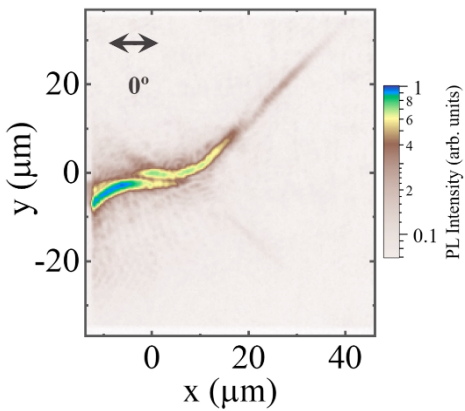

**Video S1**. Maps of the polariton emission in a directional coupler characterized by $L = 10$ μm, w = 2 μm and $d = 0.2$ μm analyzed into its linear components for different analyzer angles ranging from $\theta_d = 0°$ to $180°$.

## II. Polarization beatings at the output terminal: additional experiments.

Figure S1 compiles additional experiments, in different waveguides as those discussed in the main text, to illustrate both: a) the reproducibility of the oscillations observed at the output terminal for polaritons polarized horizontally ($\theta_d = 0°$) and vertically ($\theta_d = 90°$) and b) its independence on the polarization of the excitation.

The waveguide's parameters for these additional experiments are: $L = 10$ μm, $w = 2$ μm, and $d = 0.2$ μm. The sample is kept at 16 K and non-resonantly excited at 1.664 eV with an excitation density of 15 kW/cm$^2$.

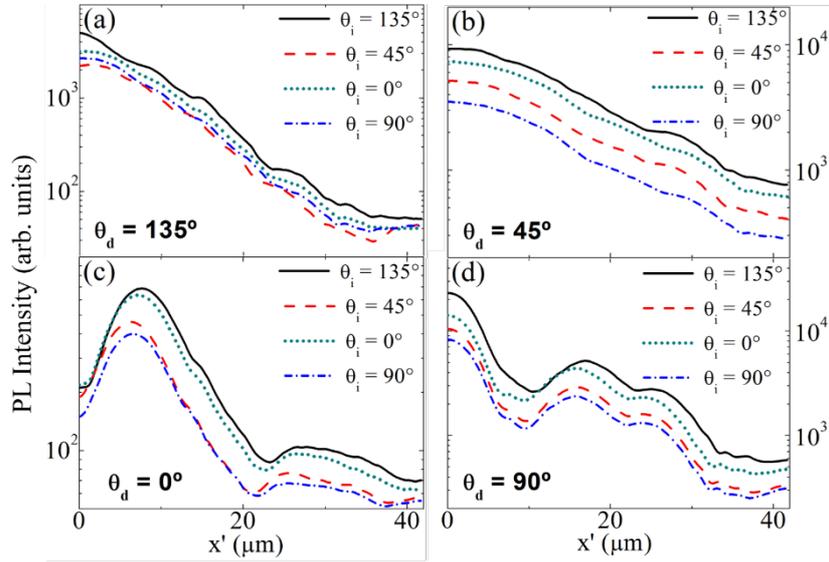

**Figure S1**. PL intensity, integrated along the transverse coordinate in the output terminal of the coupler, vs. distance, measured along the longitudinal axis at this terminal, in semilogarithmic scale, for excitation with different linearly polarized beams: $\theta_i = 135°$ (black, solid line), $\theta_i = 45°$ (red, dashed line), $\theta_i = 0°$ (dark cyan, dotted line) and $\theta_i = 90°$ (blue, dash-dot line). The four panels correspond to different analyzer angles: (a) $\theta_d = 135°$, (b) $\theta_d = 45°$, (c) $\theta_d = 0°$ and (d) $\theta_d = 90°$.

## III. Simulations of the polariton dynamics.

Here we theoretically consider in detail the polaritons polarization dynamics. For the sake of mathematical convenience, we rewrite Eqs. (1) & (2) in dimensionless units. We normalize time by the dissipation rate, $\gamma_{\text{polariton}}$, of the polaritons in the center of the waveguide, $t_{\text{normalized}} = \gamma_{\text{polariton}} t$, and the coordinate as $x_{\text{normalized}} = \sqrt{\dfrac{2D}{\hbar \gamma_{\text{polariton}}}} x$, where $D = \frac{\hbar^2}{2m_{eff}}$. For $\gamma_{polariton} = 0.24$ ps$^{-1}$ and $D = 0.55$ meV·μm², the unit of normalized time and of spatial coordinate correspond to 4.17 ps and to 2.64 μm, respectively (the unit of the dimensionless frequency is 0.158 meV).

The dimensionless equations for the order parameter functions $\psi_{r,l}$ characterizing the right- and left-handed circularly polarized polariton condensates

$$\partial_t \psi_{r,l} = (\eta_{r,l} - \gamma)\psi_{r,l} + i\left(U(x,y) + g|\psi_{r,l}|^2 + \tilde{g}|\psi_{l,r}|^2 + g_R \eta_{r,l} + \tilde{g}_R \eta_{l,r}\right)\psi_{r,l} - \frac{i}{2}\nabla^2 \psi_{r,l} + \beta(\partial_x \pm i\partial_y)^2 \psi_{l,r} + a_{r,l}(x,y,t), \qquad (S1)$$

are coupled to the equations for the reservoirs of the incoherent excitons of different spins

$$\partial_t \eta_{r,l} = -\left(\Gamma + \rho|\psi_{r,l}|^2\right)\eta_{r,l} + p_{r,l}(x,y,t). \qquad (S2)$$

In these new equations all coefficients are dimensionless (normalized): $\eta_{r,l}$ are the densities of the incoherent excitons of different spins, $\gamma$ is the coordinate dependent losses of the coherent polaritons, $U$ is the coordinate dependent effective potential for the polaritons, the nonlinear corrections to the effective potential due to polaritons and incoherent excitons of the same (orthogonal) polarization are accounted by $g(\tilde{g})$ and $g_R(\tilde{g}_R)$, $\beta$ defines the strength of TE-TM splitting (spin-orbit coupling), $a_{r,l}$ is the direct optical pump of the polaritons, $\Gamma$ is the linear losses in the exciton subsystem, $\rho$ is the exciton depletion rate due to condensation into polaritons, and $p_{r,l}$ is the optical pump intensity exciting the exciton bath. For our simulations we normalize the field $\Psi$ to make $g = 1$ and take typical values of the parameters, $g_R = 5.46$, $\tilde{g} = -0.1$, $\tilde{g}_R = -0.546$, $\beta = 0.05$, $\Gamma = 3$ and $r = 1.1$. The confining potential is rectangular with a width $w_0 = 0.54$ and depth $U_0 = 42$, outside the waveguide the polariton losses are 5 times higher than inside. In the coupling region the height of potential barrier separating the waveguides is lowered to $U_{barrier} = 27$, the distance between the waveguides, 0.076, corresponds to the experimental value of 0.2 μm.

### III.1. Excitation by a coherent pump.

#### III.1.1. Neglecting TE-TM splitting.

Let us start our analysis considering the simplest case of propagation of polaritons in a straight waveguide, what renders easy to understand results to explain the main experimental features. To have a better control over the propagating polaritons, we use a coherent pump allowing us to excite a mode of our choice. We restrict ourselves to the linear regime, obtained under low pump power densities. In the lossless limit, and neglecting the TE-TM splitting, the polariton dispersion relation is given by

$$\omega_m = \frac{1}{2}k^2 + \Omega_m, \qquad (S3)$$

Where $k$ is the wavevector along the waveguide, $\Omega_m$ are the discrete eigenvalues of the operator $\hat{L} = -\frac{1}{2}\nabla^2 - W_0(y)$, with $W_0$ the transverse confinement potential of the waveguide, and $m$ enumerate the modes, with $m = 1$ corresponding to the fundamental mode. For a rectangular potential of infinite depth and width $w_{wg}$, the cut-off frequency is given by well-known formula $\Omega_m = \frac{1}{2}\left(\frac{\pi m}{w_{wg}}\right)^2$ and the eigenfunctions are proportional to $\sin\left(\frac{m\pi y}{2w_{wg}} + \frac{\pi m}{2}\right)$. Let us remark that the frequencies are defined as the detuning from the cut-off frequency of a waveguide of infinite width.

To check how accurately Eq. (S3) describes the dispersion characteristics of the guided polaritons we performed numerical simulations of Eqs. (S1)-(S2) and found out that Eq. (S3) is very precise, especially for the fundamental mode.

#### III.1.2. Including TE-TM splitting.

In agreement with experimental findings [1], a simple perturbation analysis shows that TE-TM splitting lifts the degeneracy and each mode splits into two modes linearly polarized parallel and perpendicular to the waveguide axis [2]. For the case of the rectangular waveguide of infinite depth and width $w_{wg}$ the dispersion of the modes can then be written as

$$\omega_{m\,x,y} = \left(\frac{1}{2} \pm \beta\right)k^2 + \left(\frac{1}{2} \mp \beta\right)\left(\frac{\pi m}{w_{wg}}\right)^2, \qquad (S4)$$

where the subindices $x, y$ denote the directions of the polarization (the waveguide axis is oriented along x). To generalize this expression to a finite depth waveguide, one needs to consider the waveguide cut-off frequencies and to calculate the overlap integrals, using the waveguide eigenmodes. However, for the main conclusions discussed in this paper, it is sufficient to work with the modes of an infinite depth rectangular effective potential.

### III.1.2.1. Horizontal and vertical excitation.

To confirm that the linearly polarized modes are the eigenmodes of the waveguides, we performed numerical simulations of the polaritonic waveguide excited at $x = -5$ μm by a coherent optical pump with dimensionless frequency $\omega_{pump} = 32$, and a Gaussian spatial profile $\exp\left(-\frac{x^2+y^2}{R_{ch}^2}\right)$, $R_{ch} = 0.25$. The results are presented in Figure S2. The propagation of the polaritons excited by a linearly x-polarized beam is illustrated via the PL intensity analyzed with a polarizer oriented along (panel a$_1$) and across (panel a$_2$) the waveguide axis. It is evident that nearly all energy is confined in the x-polarization and no beating between the polarization is seen. Let us remark that the short period fringes at the ends of the waveguide originate from the reflection of the polaritons at the waveguide edges [3]. Panels (b) depict the corresponding results for linearly y-polarized excitation, rendering similar results but in this case the highest intensity is observed for y-polarized propagating polaritons. Thus, we can conclude that the eigenmodes of the polariton waveguide are linearly polarized along and perpendicular to the waveguide axis.

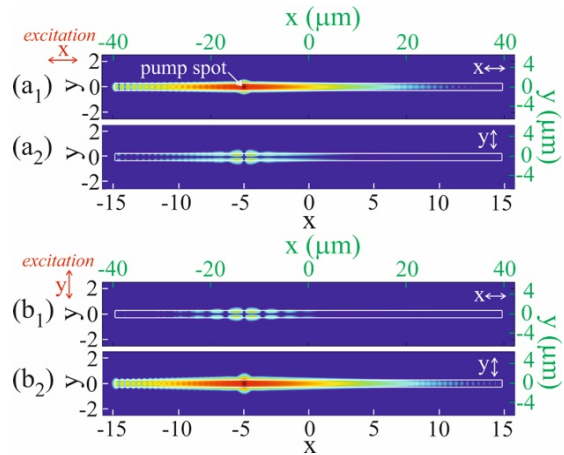

**Figure S2.** Simulated polariton distributions generated by a spatially localized coherent cw pump. The two upper panels are obtained for excitation with a x-polarized ($\theta_i = 0°$ ↔) coherent pump. Panel (a$_1$/a$_2$) shows the x/y-polarized polariton distribution. Corresponding distributions for a y-polarized ($\theta_i = 90°$ ↕) excitation are depicted in the two lower panels (b$_1$) and (b$_2$), respectively. The waveguide borders are shown by a thin white line. The abscissas and ordinates are shown in dimensionless (physical) units at the bottom (top) and left (right) axes, respectively.

### III.1.2.2. Circular and diagonal excitation.

However, a very different dynamics can be observed when the waveguides are excited by light encompassing both x and y linear polarizations. We present the results for pumps circularly polarized and linearly polarized at 45º to the waveguide axis. Let us start the discussion with circularly polarized light. As seen in Figure S3, circularly polarized light excite in equal extent both x and y polarizations [see panels ($a_1$) and ($a_3$)] and, as before, the intensities for these waveguide eigenstates do not show any oscillations. However, when the PL is analyzed by linear polarizers oriented at some angle to the waveguide axis, intensity oscillations appear, as evidenced in panels ($a_2$) and ($a_4$) for to analyzers oriented at 45º and 135º to the waveguide axis, correspondingly. A very similar dynamics is obtained for excitation by a linearly polarized pump oriented at 45º to the waveguide axis, as illustrated in panels (b1) – (b4).

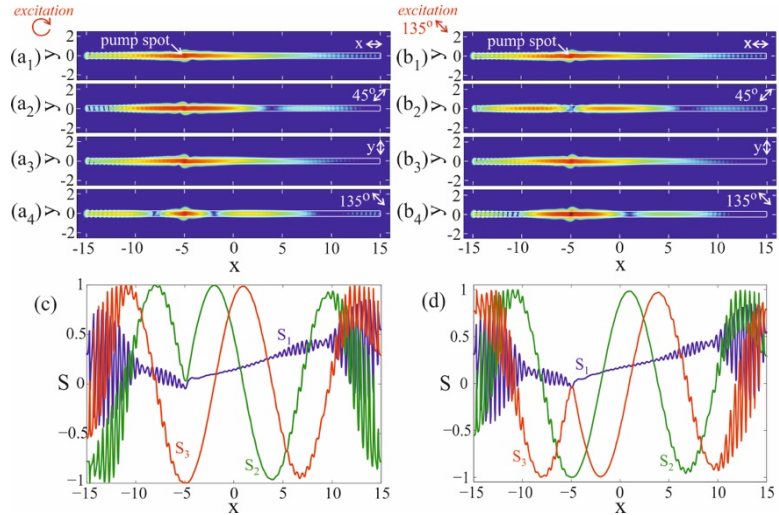

**Figure S3**. Simulated polariton distributions generated by a spatially localized coherent cw pump. Panels (a) show the polariton distributions for different angles of the linear-polarization analyzer: ($a_1$) x-polarized ($\theta_d$= 0° ↔), ($a_2$) $\theta_d$= +45°, ($a_3$) y-polarized ($\theta_d$= 90° ↕) and ($a_4$) $\theta_d$= +135°. The system is pumped by a spatially localized cw monochromatic, circularly polarized light. Corresponding distributions for a $\theta_i$ = 135° excitation are depicted in panels ($b_1$)-($b_4$). The x-dependence of the normalized Stokes parameters $S_1$(blue lines), $S_2$ (green lines) and $S_3$ (red lines) calculated along the axis of the waveguide are shown in panels (c) and (d) for circular and linear polarized excitation, respectively.

To shed more light on the polarization dynamics it is instructive to plot the dependence of the normalized Stokes parameters on the propagation distance. The spatial profiles of $S_1$, $S_2$ and $S_3$ calculated for the field at the axis of the waveguide are plotted in panels (c) and (d). The short period oscillations are due to the interference with the polaritons reflected from the waveguide edges [3]. The long period deep oscillations are of different nature as they are caused by TE-TM splitting.

As one can easily see in Fig. S3 panels (a) and (b), apart from the short period oscillations, the intensity in x and y polarizations decays monotonically, however, the PL unveils pronounced oscillations when analyzed into its 45º and 135º components. This explains the nearly perfect sinusoidal x-dependence of $S_2$. In a non-dissipating system, the position of the intensity minima in one polarization must coincide with the position of the maxima in its orthogonal polarization. However, in the presence of losses this is not the case and the minima / maxima intensities in ($a_2$, $b_2$/$a_4$, $b_4$) are noticeably shifted with respect to each other.

We shall demonstrate now that the long period polarization beating is due to the different dispersions of the x- and y-polarized modes, yielding different wavevectors for the same frequency. The dispersion relations for the fundamental and the second order modes calculated by Eq. (S3) are plotted in Figure S4(a). Panels (b) compile full k-space maps for polaritons excited by x- ($b_1$), y- ($b_2$) and circularly- ($b_3$) polarized light. For x- and y-polarized excitation only one mode is obtained; however, for a circularly polarized pump both fundamental modes are excited and thus the beating of these modes takes place. For $\omega_{pump} = 32$, the wavevectors are $k_\leftrightarrow = 6.73$ for the x-polarized fundamental mode and $k_\updownarrow = 7.2$ for the y-polarized one. This gives $\Delta k = k_\updownarrow - k_\leftrightarrow = 0.47$ and a beating period $L_{beating} = 2\pi/\Delta k = 13.3$ (35.1 μm). This result is in good agreement with the oscillation period of $S_2$ 11.9 (31.4 μm) observed in Figs. S3(c) & (d). Thus, we can identify the difference of the wavevectors for differently polarized fundamental modes as the origin of the polarization beating.

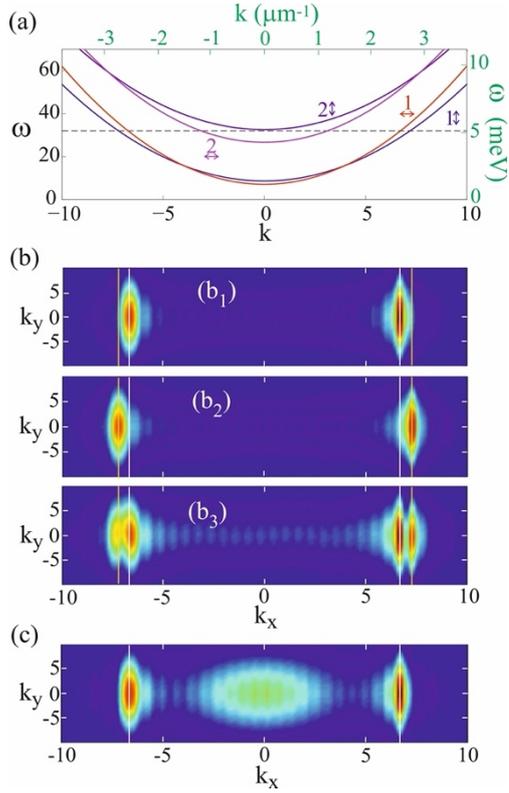

**Figure S4.** (a) Simulated polariton dispersion relations in the presence of TE-TM. The abscissas and ordinates are shown in dimensionless (physical) units at the bottom (top) and left (right) axes, respectively. The dispersion of the fundamental x(y)-polarized mode is marked as 1↔(↕); that of the second order mode is marked correspondingly as 2↔(↕). (b) Two-dimensional reciprocal-space images obtained from numerical simulations, exciting with a coherent cw pump, whose energy is marked with a horizontal dashed line in panel (a). Panel ($b_1$)/($b_2$) corresponds to x ($\theta_i = 0°$ ↔) /y ($\theta_i = 90°$ ↕) -polarized excitation, while panel ($b_3$) is for circularly polarized excitation. The vertical thin lines are guides to the eye to link the bright spectral patterns with the dispersions of differently polarized fundamental modes. Panel (c) shows the image exciting with an incoherent pump.

It should also be mentioned that TE-TM splitting also affect the period of the fringes appearing due to the reflection at the edges. The calculations based on Eq.(S4) give a period of 0.47/0.44 (1.23/1.15 μm) for the x-/y-polarization which is in full agreement with the data extracted from the direct numerical simulations [Eqs. (S1)-(S2)].

Let us remark that the decay length (defined as the propagation distance at which the intensity drops down by factor $e$) is different for the x- and y-polarized polaritons. From our numerical simulations, we extract a decay length of 2.91/2.48 (7.68/6.54 μm) for x/y polarization. The different mode widths is one of the factors contributing to the difference in the decay-lengths, with more loosely localized modes feeling higher dissipation outside the waveguides. However, the dominating factor explaining the difference in the decay lengths is the different group velocities of the modes. From the dispersion, one observes that x/y-polarized polaritons have k-vectors equal to 7.2/6.73. The ratio of the group velocities is $\frac{(0.5+\beta)k_\leftrightarrow}{(0.5-\beta)k_\updownarrow} = 1.14$ which is close to the ration of the decay lengths, 1.17.

This difference in the decay lengths explains the growth of $S_1$ for x> -5 μm in Figs. S3 (c) & (d). This implies that the amplitude of the long-period beating decreases with the propagation distance and that at large propagation distances only x-polarized polaritons will be seen. In fact, a slight decrease of the modulation is noticeable in the numerical simulations for the $S_2$ and $S_3$ Stokes parameters. This effect is analogous to the decay of the beatings of the polaritons between neighboring waveguides discussed in in Ref. [4].

### III.1.2.3. Mode structure across the waveguide.

An additional interesting fact is that the numerical simulations show that the polarization of the fundamental eigenmode is not linear at all points across the waveguide, see Fig. S2. As easily deduced from a second order perturbation analysis, if a mode is x/y-polarized in first order, the second order correction is y/x-polarized and has different parity. The structure of the fields, obtained from the results plotted in Fig. S2(a) and (b) at x=0, is shown in Fig. S5(a/b) for x/y- polarized excitation and analyzed into its x/y-polarized components (black/ blue lines), where a small signal with orthogonal polarization to the main one is clearly seen. As a consequence, a symmetric pump centered on the axis of the waveguide can excite a second order eigenmode, since the latter has a small even component in the polarization orthogonal to the polarization of the odd component. Additionally, this second order mode can also be excited if the pump is slightly displaced from the waveguide axis. The excitation of this second mode is weak but it is clearly seen in our simulations. The dependence on x of the maximum intensity, which occurs at y=0/0.2, for x/y polarized-emission, is shown in Fig. S5(c) for x-polarized excitation. The equivalent dependence for y-polarized excitation is depicted in Fig. S5(d). At large x, both intensities decay with the same rate, attesting that they belong to the same mode. However, around x~ -5 the y/x polarized intensity in (c/d) shows conspicuous oscillations, arising from the interference of the fundamental and second order modes. These oscillations extend up to x ~ 0 in panel (d), attesting that some contribution from the second order mode is still present in the field structure depicted in panel (b).

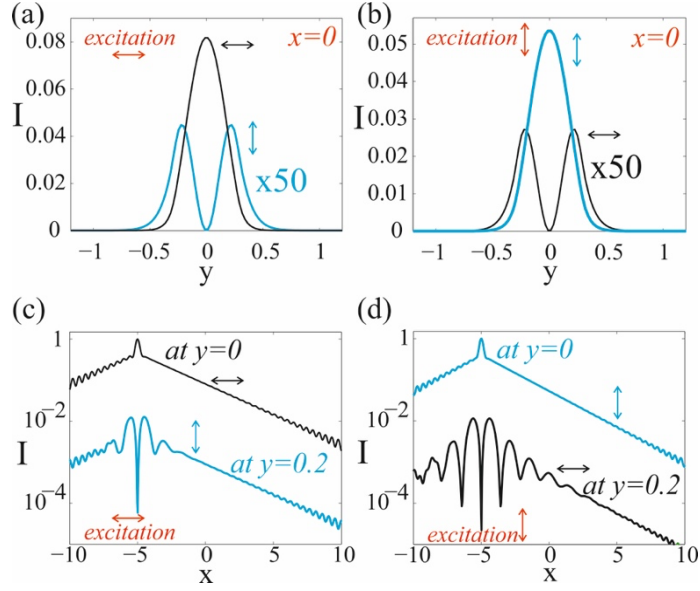

**Figure S5**. Simulated spatial intensity distributions of the x- ($\theta_d = 0°$ ↔, black lines) and y-polarized ($\theta_d = 90°$ ↕, blue lines) components of the polariton field. Panel (a)/(b) depicts the intensities along y (i.e., across the waveguide axis) at x = 0 for ($\theta_i = 0°$ ↔) / ($\theta_i = 90°$ ↕) -polarized coherent excitation: the intensity of the y/x-polarized signal is magnified by factor of 50. Panel (c)/(d) compiles the intensities along x (i.e., along the waveguide axis) at y = 0 for ($\theta_i = 0°$ ↔) / ($\theta_i = 90°$ ↕) -polarized coherent excitation.

### III.2. Excitation by an incoherent pump.

Now, we turn to the study of the polariton dynamics pumped by an incoherent pump, creating a bath of incoherent excitons. This excitation disregards the term $a_{r,l}$ in Eq. (S1) and considers only $p_{r,l}$ in Eq. (S2). For our numerical simulations, we use the super-Gaussian function $p_{r,l} = p_0 \exp\left(-\frac{(x^2+y^2)^4}{w_{ip}^8}\right)$, where $p_0$ is the intensity of the pump and $w_{ip}$ is its width. We use $w_{ip} = 0.9$ and $p_0 = 35$, well above the polariton condensation threshold. A weak random noise in the polariton field was also taken as initial conditions. To allow a direct comparison between the coherent and incoherent pump scenarios, the parameters were chosen so that the frequency of the excited polaritons is the same for both cases. The two-dimensional reciprocal-space emission image, exciting with an incoherent pump, is shown in Fig. S4 (c): it is clearly seen that the wavevector, $k_x$, of the propagating polaritons coincides with that found for coherent excitation.

The results of our simulations in the straight waveguide are presented in Fig. S6 (a), showing the distribution of x- and y-polarized polaritons for the stationary solution. It is clearly distinguished that the excited mode is x-polarized, however, as discussed above, the mode that has a dominating x-polarization of even parity also possesses a weak y-polarized component of odd parity. Thus, the polaritons are purely x-polarized only at the axis of the waveguide. Out-of-axis, their polarization is slightly elliptical, as borne out by the weak maxima of the y-polarized intensity close to the waveguide borders. In this case no beating in the y-polarization is seen (apart from the fringes caused by the reflection of the polaritons at the waveguide edges). This means that the incoherent pump excites only the fundamental mode with dominating x-polarization. The dispersion relation yields that this mode is the lowest in energy, therefore, for the chosen parameters, polariton condensation occurs at the thermodynamically favorable mode, in agreement with our experiments and with a previous theoretical analysis [5].

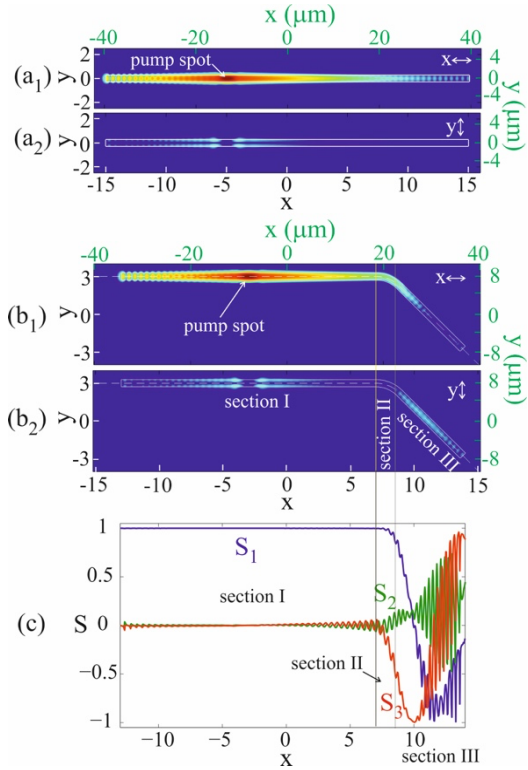

**Figure S6**. Simulated polarized polariton distributions generated by an incoherent x-linearly-polarized pump analyzed into its x ($\theta_d = 0°$ ↔) and y ($\theta_d = 90°$ ↕) – polarized components. Panels (a) illustrates the case of a straight waveguide and panels (b) that of a waveguide consisting of two straight parts (sections I and III) connected by a segment of an annular waveguide (section II). The boundaries of the waveguides are marked by thin white lines. The abscissas and ordinates are shown in dimensionless (physical) units at the bottom (top) and left (right) axes, respectively. Panel (c) shows the normalized Stokes parameters $S_1$ (blue lines), $S_2$ (green lines) and $S_3$ (red lines) as functions of x for the waveguide considered in panels (b).

Having shown that in a straight waveguide an incoherent pump excites only one mode, and thus no polarization oscillations are possible, to explain the experimental results in our couplers, it is essential to simulate a waveguide which changes its direction. The profile used for the simulations is presented in Fig. S6 (b): it consists of a horizontal part (section I) that is connected to another part of a straight waveguide (section III) by a segment of an annular waveguide (section II). A x-polarized polariton condensate is created in section I via an incoherent pump. Its propagation does not show any polarization oscillations, as evidenced by the Stokes parameters plotted in Fig. S6 (c), where the weak frequent oscillations appear because of waves reflected from section II and the edges of the waveguide. However, as the polaritons propagate, the x-polarized wave ceases to be an eigenmode in section II: the polaritons start to redistribute over different horizontal (x↔) and vertical (y↕) polarizations as demonstrated by the continuous changes of the Stokes parameters.

In section III, the waveguide is oriented at 45º to the x axis and the eigenmodes are linearly polarized, but with polarizations oriented at 45º and 135º degrees to this axis. Because of the just mentioned polarization redistribution in section II, in general, in section III, the polaritons polarization is given by a linear combination these eigenstates. These modes have different wavevectors and therefore the density of polaritons oscillates between x and y polarizations, as discussed in Section III.1.2.2. It is noticeable in Fig. S6 (b) that x-polarization gets depleted while polaritons appear in y-polarization. This explains satisfactorily the experimentally observed polarization oscillations in the output terminal.

**References**

[1]     Dasbach, G.; Diederichs, C.; Tignon, J.; Ciuti, C.; Roussignol, Ph.; Delalande, C.; Bayer, M.; Forchel, A. Polarization selective polariton oscillation in quasi-onedimensional microcavities. *Phys. Status Solidi C* **2005**, *2*, 779-782.

[2]     Shelykh, I.A.; Nalitov, A.V.; Iorsh, I.V. Optical analog of Rashba spin-orbit interaction in asymmetric polariton waveguides. *Phys. Rev. B* **2018**, *98*, 155428.

[3]     Antón, C.; Liew, T.C.H.; Tosi, G.; Martín, M. D.; Gao, T.; Hatzopoulos, Z.; Eldridge, P. S.; Savvidis, P.G.; Viña, L. Energy relaxation of exciton-polariton condensates in quasi-one-dimensional microcavities. *Phys. Rev. B* **2013**, *88*, 035313.

[4]     Beierlein, J.; Rozas, E.; Egorov, O.A.; Klaas, M.; Yulin, A.; Suchomel, H.; Harder, T.H.; Emmerling, M.; Martín, M.D.; Shelykh, I.A.; Schneider, C.; Peschel, U.; Viña, L.; Hoffling, S.; Klembt, S. Propagative oscillations in codirectional polariton waveguide couplers. *Phys. Rev. Lett.* **2021**, *126*, 075302.

[5]     Malpuech, G.; Glazov, M.M.; Shelykh, I.A.; Bigenwald, P.; Kavokin, K.V. Electronic control of the polarization of light emitted by polariton lasers. *Appl. Phys. Lett.* **2006**, *88*, 111118.